\documentclass[fleqn,usenatbib]{mnras}
\usepackage{newtxtext,newtxmath}

\usepackage[T1]{fontenc}
\usepackage{ae,aecompl}

\usepackage{graphicx}	
\usepackage{amsmath}	
\usepackage{amssymb}	

\newcommand{\msun}{\rm\, M_\odot}

\title[AGN in dwarf galaxies]{Difficulties in Mid-Infrared selection of AGN in dwarf galaxies}

\author[A. Lupi et al.]{
Alessandro Lupi,$^{1}$\thanks{E-mail: alessandro.lupi@sns.it}
Tullia Sbarrato,$^{2}$ and
Stefano Carniani$^{1}$
\\
$^{1}$Scuola Normale Superiore, Piazza dei Cavalieri 7, Pisa IT-56126, Italy\\
$^{2}$Dipartimento di Fisica, Universit\`{a} degli Studi di Milano-Bicocca, Piazza della Scienza 2, Milano IT-20123, Italy\\
}

\date{Accepted XXX. Received YYY; in original form ZZZ}

\pubyear{2020}

\begin{document}
\label{firstpage}
\pagerange{\pageref{firstpage}--\pageref{lastpage}}
\maketitle

\begin{abstract}
While massive black holes (MBHs) are known to inhabit all massive galaxies, their ubiquitous presence in dwarf galaxies has not been confirmed yet, with only a limited number of sources detected so far. Recently, some studies proposed infrared emission as an alternative way to identify MBHs in dwarfs, based on a similar approach usually applied to quasars. In this study, by accurately combining optical and infrared data taking into account resolution effects and source overlapping, we investigate in detail the possible limitations of this approach with current ground-based facilities, finding a quite low ($\sim$0.4 per cent) fraction of active MBH in dwarfs that are luminous in mid-infrared, consistent with several previous results. Our results suggest that the infrared selection is strongly affected by several limitations that make the identification of MBHs in dwarf galaxies currently prohibitive, especially because of the very poor resolution compared to optical surveys, and the likely contamination by nearby sources, although we find a few good candidates worth further follow-ups. Optical, X-ray and radio observations, therefore, still represent the most secure way to search for MBH in dwarfs.
\end{abstract}




\section{Introduction}
Massive black holes (MBHs) are ubiquitous in the Universe, and inhabit all massive galaxies \citep[e.g.][]{ferrarese00}. MBHs are typically observed via their accretion-powered radiation as Active Galactic Nuclei (AGN), whose impact on to the host galaxy is invoked to explain massive galaxy quenching \citep[e.g.][]{silk98,dimatteo05} 
and the emergence of the MBH-galaxy correlations \citep{gultekin09,kormendy13bh}. However, the demographics of MBHs are not yet well constrained \citep[see, e.g.][for reviews]{reines16,greene19} and the role of AGN feedback in dwarf galaxies is just starting to be explored \citep[][]{penny18,mackaydickey19,manzanoking19}. In particular, theoretical studies have shown that the MBH-halo occupation fraction can be low in dwarf galaxies \citep[e.g.][]{greene12,miller15,habouzit17}, and even in systems hosting MBHs, their growth is strongly suppressed because of the typically low gas densities in the host (compared to more massive disc galaxies) and the strong impact of supernova explosions, both at low- and high-redshift \citep{dubois14,anglesalcazar17FIRE,prieto17,trebitsch18}. Observationally, this could be partially reflected in MBHs in dwarfs being inactive for most of their life, hence in the small number of low-luminosity AGN found \citep{reines13,mezcua16,mezcua18,reines19}, although there are also strong observational biases that could make these AGN difficult to find. On the other hand, if MBHs were present in many dwarfs, and efficiently grew, their feedback could significantly affect their host \citep[see, e.g.][]{penny18,mackaydickey19,manzanoking19}, could play a role in the reionisation of the Universe \citep{volonterignedin09}, could help removing gas from massive disc progenitors \citep[e.g.][]{peirani12}, and also mitigate the `too-big-to-fail' problem \citep{garrisonkimmel13}.
Recently, \citet[K19 hereafter;][]{kaviraj19} tried to better assess the role of AGN feedback in dwarfs by jointly analysing the Hyper-Supreme Cam Subaru Strategic Program \citep{aihara18b} and the {\it WISE} \citep{wright10} surveys, finding that AGN in dwarfs could exhibit very large bolometric luminosities, hence they could play a significant role in their host evolution.
However, a clear consensus on the identification of AGN in dwarfs is still missing \citep[see, e.g.][]{satyapal14,sartori15,marleau17}, in particular because of the low resolution of WISE ($\sim 6\arcsec$) relative to current optical surveys, resulting in a strong source overlap, and the possible contamination of infrared emission by star-formation activity, that could mimic AGN activity \citep{hainline16,satyapal18}, especially in dwarf galaxies \citep{hainline16}.
In this Letter, we build-up on the K19 work by re-analysing their dwarf sample in more detail in the aim at better disentangling plausible AGN in dwarf from star-forming galaxies by taking into account possible source overlapping in the sample, and also assess the MBH properties. 

In Section \ref{sec:data}, we introduce the two datasets employed and the selection criteria applied, in Section \ref{sec:results} we present our results, and in Sections \ref{sec:discussion} and \ref{sec:conclusions} we discuss the caveats of the study and draw our conclusions.

\section{Data Analysis}
\label{sec:data}
\subsection{Data selection and matching}
In this work, we employ the galaxy catalog of the Hyper-Supreme Cam Subaru Strategic Program \citep[HSC-SSP hereafter;][]{aihara18b}, an imaging survey in {\it grizy} and 4 narrow-band filters with a resolution of $\sim$0.6\arcsec. In particular, we use the Wide layer Data Release 1 \citep{aihara18a}, which provides an unprecedented census of dwarf galaxies, being about 4 mag deeper  than SDSS. For our analysis, we employ the photometric redshifts ($z_{\rm best}$), stellar masses and star formation rates obtained by \citet{tanaka18} using the \textsc{mizuki} code \citep{tanaka15}.
To select AGN, as in K19, we employ the {\it WISE} survey, that mapped the sky in 4 bands ranging from 3.4 to 22 $\mu$m, with angular resolutions of $\sim6$\arcsec\ in bands 1,2,3, and 12\arcsec\ in band 4.
For our analysis, we cross-matched the catalogues using a search radius of 4\arcsec, as in K19. Here (K19 does not give details on how the matching was performed) we match each HSC-SSP source to a {\it WISE} source, when available. This choice is meant to avoid a biased identification of the sources based on the relative distance between the flux centres, which could be a severe problem if two surveys have extremely different angular resolutions, and several sources in the HSC-SSP sample overlap with a single detection in {\it WISE}.

\begin{figure}
    \centering
    \includegraphics[width=0.93\columnwidth]{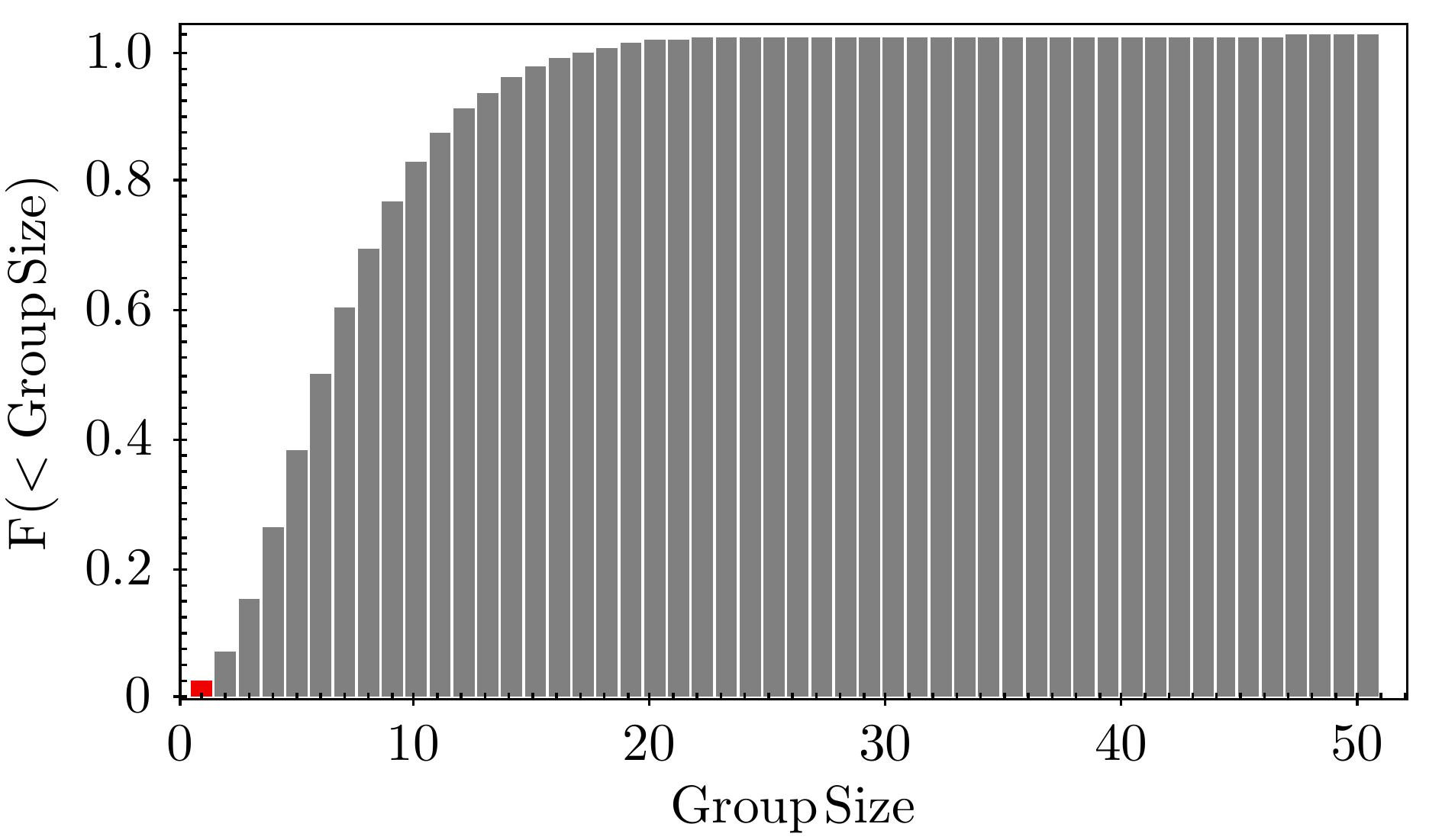}
    \caption{Normalised cumulative distribution of the group size for the cross-matched sources in our fiducial dataset. The histogram shows that most of the galaxies in the sample are grouped together, with only 2 per cent (shown in red) being isolated sources in both catalogues.}
    \label{fig:groupsize}
\end{figure}

Finally, before filtering out massive galaxies to look at dwarfs, we grouped our data catalogue according to the {\it WISE} source designation, i.e. every source in the group is associated to the same {\it WISE} source, \footnote{As an additional check, we repeated the grouping using a distance criterion with radius 6\arcsec\ (consistent with the best angular resolution of WISE), finding similar results to our fiducial grouping method.} and estimate the minimum and maximum stellar masses in each group. The cumulative distribution of group sizes in our fiducial matched dataset is shown in Fig.~\ref{fig:groupsize}, and highlights that most of the HSC-SSP sources (98 per cent) belong to groups with at least 2 members, i.e. multiple HSC-SSP sources are assigned to the same {\it WISE} source. 

\subsection{The dwarf galaxy sub-sample}
To identify dwarf galaxies in the sample, we first apply the same selection criteria as in K19, i.e. the confidence in the redshift estimation by \textsc{mizuki} $z_{\rm conf,best}>0.8$, signal-to-noise (S/N) ratio in {\it WISE} bands 1 (W1) and 2 (W2) larger than 5, stellar masses 
$M_{\rm star}<10^9\msun$, and $0.1<z_{\rm best}<0.3$. Then, we further refine our dataset by requiring $m_{\rm W1}-m_{\rm W2}>0.52$ \citep{satyapal14}, as in K19, in order to identify potential AGN in dwarf galaxies. Unlike K19, we also require here S/N$_{\rm W3}>$2, and we discuss the impact of this choice in appendix~\ref{app:SNcut}.
The resulting dataset consists of about 500 objects of which only 15 are not in groups. 

\section{Results}
\label{sec:results}

\subsection{Dwarf AGN candidates}

\begin{figure}
    \centering
    \includegraphics[width=0.93\columnwidth]{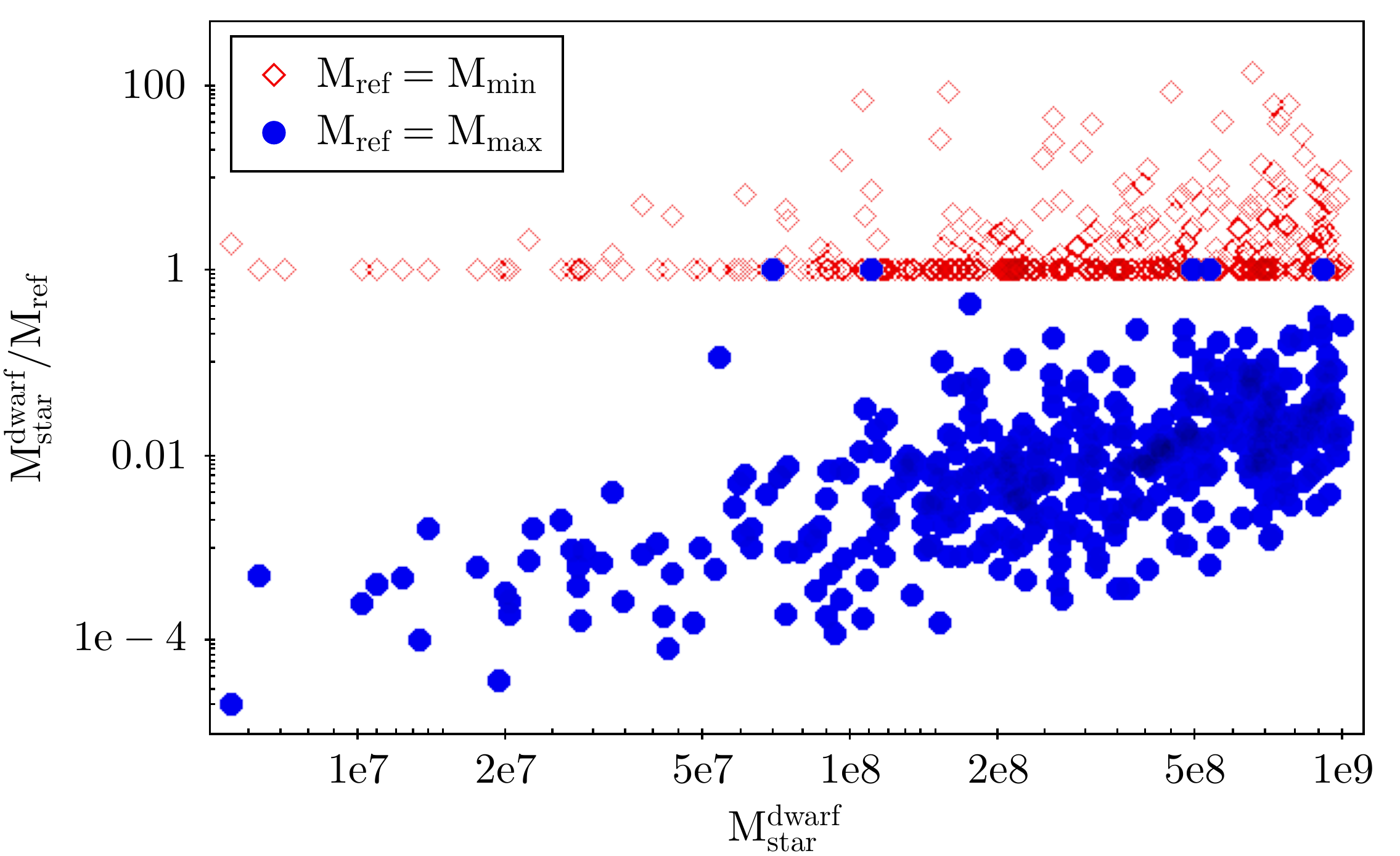}
    \caption{Stellar mass ratio as a function of the dwarf AGN candidate stellar mass for galaxies in our sample belonging to groups. The blue dots show $M_{\rm star}^{\rm dwarf}/M_{\rm star,max}$ whereas the red diamonds correspond to $M_{\rm star}^{\rm dwarf}/M_{\rm star,min}$. Our results show that most of the identified dwarf galaxies belong to groups with at least one more massive companion (blue dots), and they often represent the smallest galaxy in the group (red diamonds). This suggests that the infrared emission is likely dominated by the more massive systems in the group, or at least contaminated by it.}
    \label{fig:mmax}
\end{figure}
In Fig.~\ref{fig:mmax}, we report the ratio between the actual stellar mass of the dwarf AGN candidate and the minimum (maximum) stellar mass in the group the candidate belongs to, as red diamonds (blue dots). Our result shows that most of the galaxies in each {\it WISE} detection belong to groups where the most massive galaxy is about 100 times more massive than the selected one, corresponding to a median value of $M_{\rm star,max}\sim 5\times 10^{10}\msun$. In addition, they often represent the smallest system of the group
and this suggests that probably most of the emission detected by WISE, both in the case of an AGN or of dust-powered emission, is either coming from the most massive system in the group, expected to be more metal (and dust) rich and to possibly host a MBH, or at least contaminated by it.
\begin{figure*}
    \centering
    \includegraphics[width=0.16\textwidth]{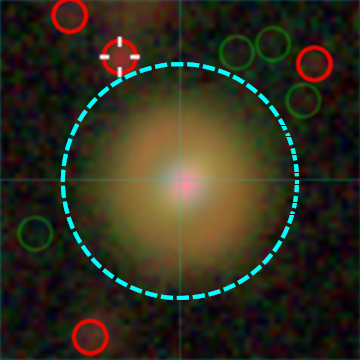}
    \includegraphics[width=0.16\textwidth]{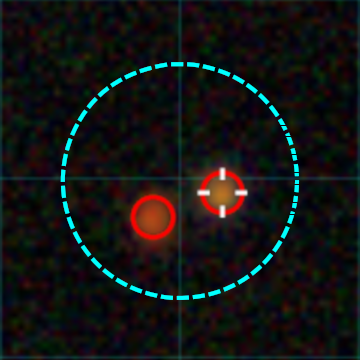}
    \includegraphics[width=0.16\textwidth]{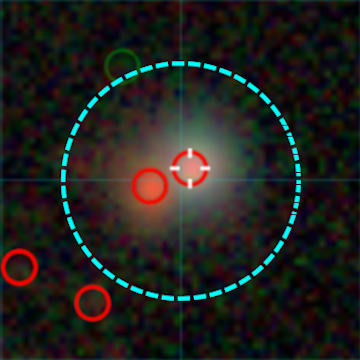}
    \includegraphics[width=0.16\textwidth]{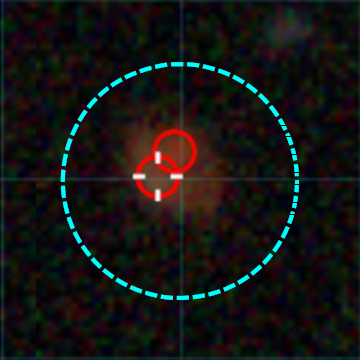}
    \includegraphics[width=0.16\textwidth]{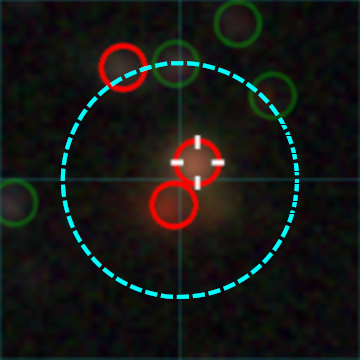}
    \caption{HSC-SSP maps of a few selected dwarf galaxies, in panels 10\arcsec\ wide centred at the coordinates of the corresponding {\it WISE} source. The dwarfs are identified by white/red target symbols, whereas the red circles correspond to galaxies with $M_{\rm star}>10^{10}\msun$. We report the {\it WISE} angular resolution in the W3 band as cyan dashed circles with diameter $\sim 6.5$\arcsec. These dwarfs probably correspond to satellites or background/foreground systems, and most likely the W3 emission is not associated with them but with the most massive nearby object.}
    \label{fig:maps}
\end{figure*}

To better highlight our results, we report in Fig.~\ref{fig:maps} a few examples of the HSC-SSP maps, centred at the location of the associated {\it WISE} source (from left to right, J090305.92+014850.8, where \textsc{mizuki} misses the central source, J223756.08+013907.5, J143926.07+000117.0, J115044.34+003912.0, and J021717.89-045408.4), where the target dwarfs are identified by the red/white target symbol, galaxies more massive than $10^{10}\msun$ are shown as red circles, and the resolution in the {\it WISE} W3 band as cyan dashed circles.

Here, we conservatively assume that only the dwarf galaxies representing the most massive object in their group could be actual AGN candidates. This further reduces our sub-sample to only 5 objects.

By adding the isolated dwarf sub-sample, we get a total of 20 possible AGN (Table~\ref{tab:objects}) out of about 5000 dwarf galaxies (with S/N$_{\rm W3}>2$), corresponding to roughly 0.4 per cent, unlike the 10 per cent found by K19.
We stress that, if a more conservative cut on S/N$_{\rm W3}>5$ is employed, the sub-sample of AGN candidates will result in only 7 objects.

We notice that, if we want to employ a two-colour criterion to identify AGN in dwarfs \citep{jarrett11}, we first need to impose S/N$_{\rm W3}>5$, reducing the sample to the same 7 objects just mentioned.

\subsection{BH mass and AGN energetics}
\begin{figure}
    \centering
    \includegraphics[width=0.93\columnwidth]{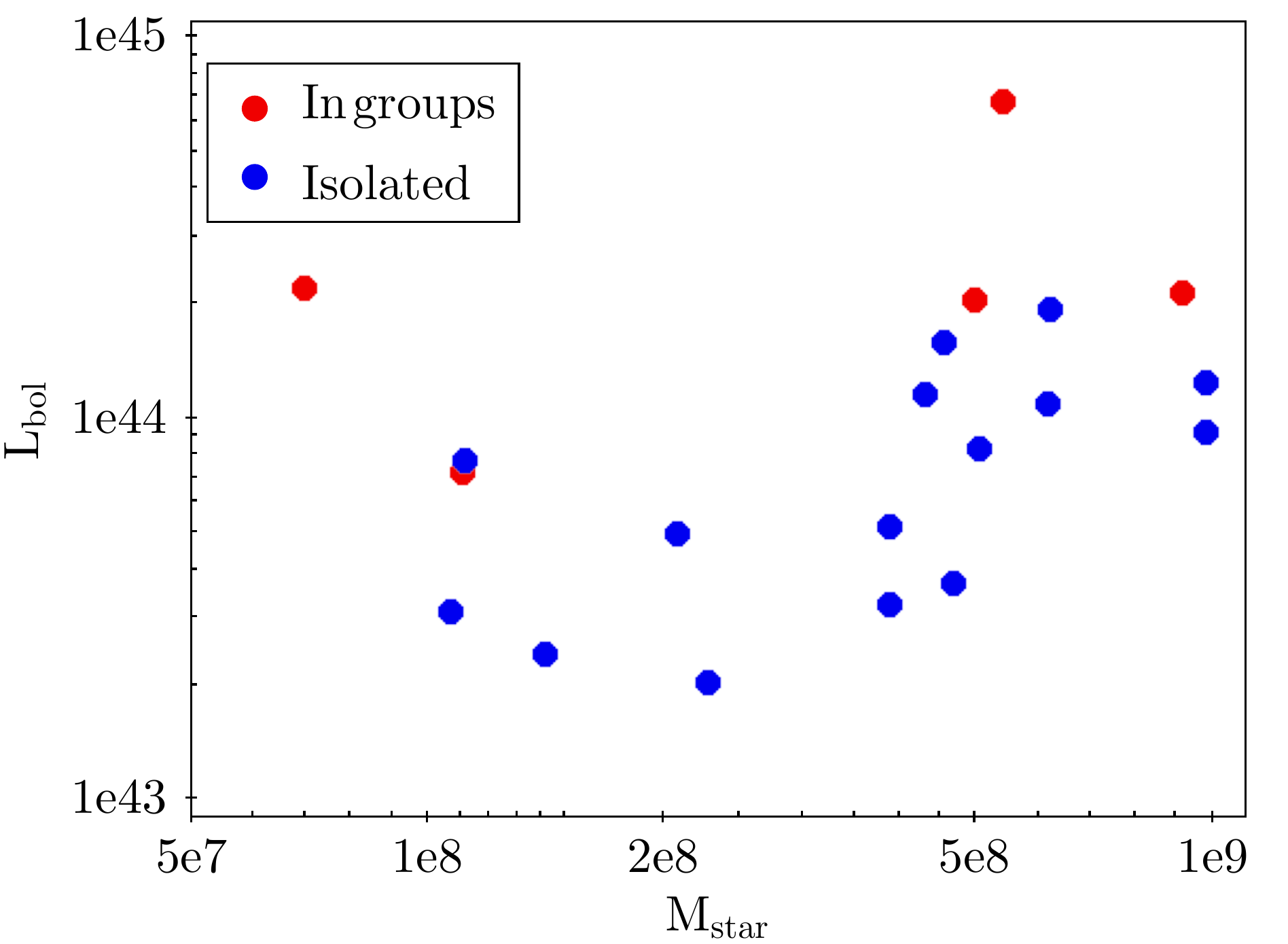}
    \caption{Bolometric luminosity of the dwarf AGN candidates assuming the \citet{richards06} bolometric correction. The blue and red dots correspond to the isolated and `in groups' sources.} 
    \label{fig:lbol}
\end{figure}

We now assess the expected AGN luminosity from our fiducial sub-sample. 
Assuming a quasar spectrum, the bolometric luminosity $L_{\rm bol}$ can be determined from the W3 flux (expected to not depend on the AGN luminosity) as $L_{\rm bol}\sim 12 L_{\rm W3}$ \citep{richards06}. 
Obviously, if the IR emission is not dominated by an AGN, this conversion would overestimate the $L_{\rm bol}$, and this could likely be the case when single colour cut are applied \citep{hainline16}.
In Fig.~\ref{fig:lbol}, we show the bolometric luminosity for our fiducial sub-sample, with dwarfs in groups shown in red and isolated ones in blue. The median bolometric luminosity is $L_{\rm bol}=10^{44}\rm\, erg\, s^{-1}$, two orders of magnitude smaller than in K19. Even if we consider the brightest AGN sample, i.e. those with S/N$_{\rm W3}>5$, as done in K19, our median bolometric luminosity does not change significantly, in clear contrast with K19, where the median value is extremely high, consistent with a MBH mass of $10^8\msun$ at best, assuming the accretion is occurring at the Eddington limit.

\begin{figure}
    \centering
    \includegraphics[width=0.93\columnwidth]{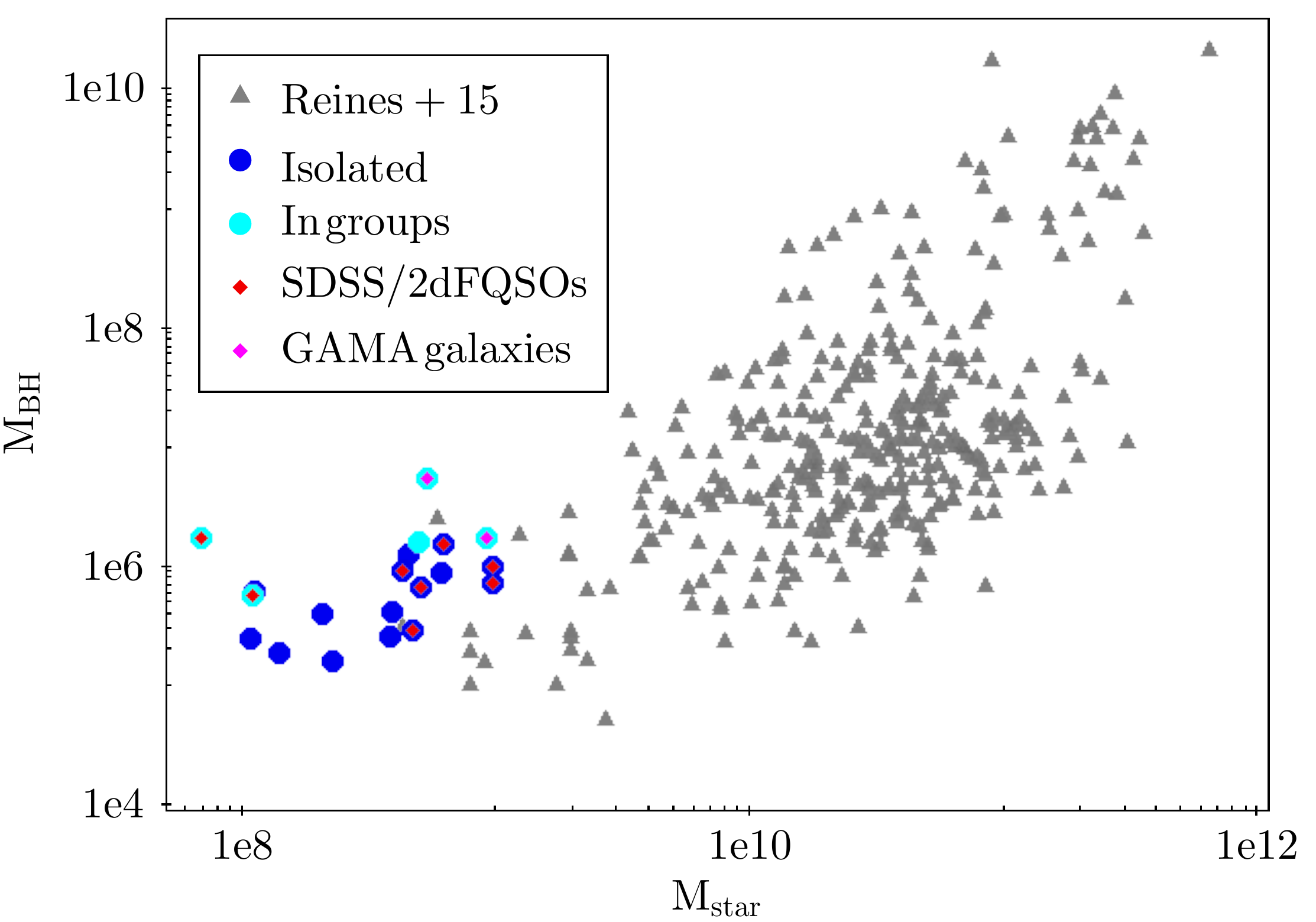}
    \caption{BH mass-stellar mass correlation for our dwarf galaxy sub-sample, obtained assuming accretion at the Eddington limit, compared with the \citet{reines15} data, shown as grey triangles (that include both AGN and quiescent MBHs). The isolated systems are shown as blue dots, whereas those in groups as cyan dots. The overalapped red and magenta diamonds correspond to the systems spectroscopically identified by SDSS/2dF as QSOs and by GAMA as star-forming galaxies, respectively, hence could have an inaccurate estimate of the stellar/MBH mass (they are almost certainly not dwarfs). Our candidates likely represent the low-end of the distribution, that seems to flatten at very-low stellar masses.}
    \label{fig:correlation}
\end{figure}

If we further assume that the observed AGN are accreting at the Eddington limit, we can compare this data with that by \citet{reines15}, to assess whether these MBHs are consistent with the local BH mass-$M_{\rm star}$ correlation. 
The results are shown in Fig.~\ref{fig:correlation}, with the cyan dots corresponding to our `in groups' sub-sample, the blue ones to the isolated sub-sample, and the red and magenta diamonds to the systems spectroscopically identified by SDSS as QSOs or by GAMA as star-forming galaxies (see below). The data by \citet{reines15} are shown as grey triangles. Our sub-sample clearly represents a natural extension of the MBH-stellar mass relation down to the dwarf galaxy regime, and exhibits a flattening to MBH masses of about $M_{\rm BH}=10^{5-6}\msun$.
If we instead conservatively assume that these MBHs are accreting at 10 per cent of the Eddington limit, these objects would move above the \citet{reines15} sample.

Thanks to the small number of objects in our sample, reported in Table~\ref{tab:objects}, we can also check our AGN candidates one by one to further confirm our results. To further confirm our analysis, we also recomputed the stellar masses in our sample using the k-corrections by \citet{chilingarian10} and the stellar mass estimates by \citet{taylor11}, finding reasonably consistent results with those by \textsc{mizuki}. We notice however that these results are valid as long as the photometric redshift estimate by \textsc{mizuki} is correct.
\begin{table*}
    \centering
   \caption{Dwarf AGN candidates in our sub-sample. The first three columns are the stellar mass, the SFR and the best photometric redshift estimated by \citet{tanaka18}, the fourth and fifth ones correspond to the RA and DEC coordinates of the source, and the sixth to the associated {\it WISE} target (the * symbol bracketing the {\it WISE} names highlights sources with a QSO/quasar identification in SDSS). The seventh column is the luminosity in the W3 band, determined employing the photometric redshift by \citet{tanaka18}, and the last column is the expected MBH mass for an Eddington-accreting MBH.}
     \begin{tabular}{c|c|c|c|c|c|c|c|c}
\hline
    $M_{\rm star}$ & $\dot{M}_{\rm star}$ & $z_{\rm best}$ & RA & DEC & {\it WISE} Name & S/N$_{\rm W3}$ & $L_{\rm W3}$ & $M_{\rm BH}$\\
    $(\msun)$ & $(\msun\, yr^{-1}$) & & & (degree) & (degree) & &($\rm erg\, s^{-1}$) & $(\msun)$\\
    \hline
  $4.975\times 10^{8}$ & $0.339$ & $0.22$ &  $29.9534$ & $-5.9557$ & J015948.97-055720.9 & 4.2 &  $1.686\times 10^{43}$ & $1.62\times 10^{6}$ \\
  $9.782\times 10^{8}$ & $0.648$ & $0.17$ &  $37.3414$ & $-4.7055$  & *J022921.94-044219.6* & 3.7 & $7.589\times 10^{42}$ & $7.29\times 10^{5}$ \\
  $3.876\times 10^{8}$ & $0.186$ & $0.11$ &  $32.7561$ & $-4.0269$  &  J021101.48-040137.0 & 3.1 & $2.692\times 10^{42}$ & $2.58\times 10^{5}$ \\
  $4.674\times 10^{8}$ & $0.126$ & $0.11$ &  $35.6559$ & $-3.8120$ &  *J022237.42-034842.8* & 3.3 & $3.063\times 10^{42}$ & $2.94\times 10^{5}$  \\
  $5.399\times 10^{8}$ & $0.648$ & $0.20$ & $182.1590$ & $-0.8007$  & J120838.02-004802.0 & 10.0 & $5.605\times 10^{43}$ & $5.38\times 10^{6}$  \\
  $2.085\times 10^{8}$ & $0.183$ & $0.14$ & $213.9344$ & $-0.1648$ & J141544.25-000953.1 & 2.7 &$4.099\times 10^{42}$ & $3.93\times 10^{5}$ \\
  $4.318\times 10^{8}$ & $0.494$ & $0.12$ & $215.1228$ & $-0.6766$ & *J142029.45-004035.7* & 7.7 & $9.497\times 10^{42}$ & $9.12\times 10^{5}$  \\ 
  $9.187\times 10^{8}$ & $0.911$ & $0.22$ & $214.8088$ & $-0.9984$ & J141913.94-005952.6 & 3.8 & $1.763\times 10^{43}$ & $1.69\times 10^{6}$ \\
  $1.111\times 10^{8}$ & $0.080$ & $0.16$ & $136.5073$ & $0.3396$ & *J090601.68+002020.6* & 2.3 & $5.969\times 10^{42}$ & $5.73\times 10^{5}$  \\
  $5.057\times 10^{8}$ & $0.592$ & $0.10$ & $177.8722$ & $1.1500$ &  *J115129.34+010859.9* & 6.7 &$6.908\times 10^{42}$ & $6.63\times 10^{5}$ \\ 
  $9.791\times 10^{8}$ & $0.247$ & $0.12$ & $179.8719$ & $1.0984$ & *J115929.23+010553.7* & 7.0 &$1.031\times 10^{43}$ & $9.90\times 10^{5}$  \\
  $6.171\times 10^{8}$ & $0.666$ & $0.17$ & $181.8775$ & $0.5800$ & J120730.58+003447.6 & 2.8 & $8.982\times 10^{42}$ & $8.62\times 10^{5}$   \\
  $3.882\times 10^{8}$ & $0.165$ & $0.15$ & $219.4741$ & $0.2469$ & J143753.75+001448.7 & 2.2 & $4.293\times 10^{42}$ & $4.12\times 10^{5}$  \\
  $6.974\times 10^{7}$ & $0.059$ & $0.23$ & $332.1132$ & $0.9429$ & *J220827.27+005636.4* & 3.2 &$1.810\times 10^{43}$ & $1.74\times 10^{6}$ \\
  $4.540\times 10^{8}$ & $0.155$ & $0.14$ & $336.9699$ & $0.7419$ & J222752.79+004431.1 & 5.9 &$1.302\times 10^{43}$ & $1.25\times 10^{6}$ \\
  $1.073\times 10^{8}$ & $0.042$ & $0.10$ & $336.7726$ & $0.8505$ & J222705.43+005101.4 & 2.4 & $2.569\times 10^{42}$ & $2.47\times 10^{5}$ \\
  $1.122\times 10^{8}$ & $0.028$ & $0.11$ & $131.1240$ & $1.7617$ & J084429.75+014541.7 & 5.2 &$6.417\times 10^{42}$ & $6.16\times 10^{5}$  \\
  $6.197\times 10^{8}$ & $0.749$ & $0.10$ & $134.6247$ & $1.8426$ & *J085829.92+015033.3* & 15.1 & $1.590\times 10^{43}$ & $1.53\times 10^{6}$  \\ 
  $1.413\times 10^{8}$ & $0.047$ & $0.10$ & $134.6318$ & $1.9469$ & J085831.57+015648.7 & 2.3 & $1.980\times 10^{42}$ & $1.90\times 10^{5}$ \\
  $2.288\times 10^{8}$ & $0.035$ & $0.10$ & $247.9543$ & $42.8779$ & J163148.98+425240.4 & 2.2 & $1.686\times 10^{42}$ & $1.62\times 10^{5}$ \\
 \hline
   \end{tabular}
    \label{tab:objects}
\end{table*}
  
\begin{itemize}
    \item {\bf J022921.94-044219.6}: based on spectroscopic identification by SDSS, this is a broad-line AGN at $z=0.777$; 
    \item {\bf J022237.42-034842.8}: based on spectroscopic identification by SDSS, this is a QSO at $z=0.996$; 
    \item {\bf J090601.68+002020.6}: the HSC-SSP source in our sub-sample has a companion, for which SDSS has both photometric and spectroscopic measures identifying it as a QSO at $z=0.701$. The same companion is identified by HSC-SSP as a dwarf galaxy with $M_{\rm star}\sim 10^7\msun$;
    \item {\bf J115929.23+010553.7}: based on spectroscopic identification by SDSS, this is a QSO at $z=0.667$; 
    \item {\bf J220827.27+005636.4}: based on spectroscopic identification by SDSS, this is a QSO at $z=2.235$; 
    \item {\bf J085829.92+015033.3}: based on spectroscopic identification by SDSS, this is a broad-line AGN at $z=0.629$; 
    \item {\bf J142029.45-004035.7}: quasar identification from the 2dF survey \citep{croom04}, \footnote{The 2dF QSO Redshift Survey (2QZ) was compiled by the 2QZ survey team from observations made with the 2-degree Field on the Anglo-Australian Telescope.} with spectroscopic redshift $z_{\rm spec}=0.71$; 
    \item {\bf J115129.34+010859.9}: quasar identification from the 2dF survey, with spectroscopic redshift $z_{\rm spec}=0.670$. 
\end{itemize}
For the aforementioned objects (highlighted in Table~\ref{tab:objects} via a * symbol bracketing their names), we expect the stellar/BH properties estimated by HSC-SSP+{\it WISE} not to be fully reliable, because of the discrepancy with the spectroscopic measurements by SDSS and 2dF.

For the following two sources, instead, we found corresponding sources at about 2\arcsec\ in the GAMA \citep{driver09,hopkins13gama}\footnote{GAMA is a joint European-Australasian project based around a spectroscopic campaign using the Anglo-Australian Telescope. GAMA input catalogue is based on data taken from the Sloan Digital Sky Survey and the UKIRT Infrared Deep Sky Survey.} spectroscopic catalogue, that suggest caution in their interpretation as AGN, although we cannot totally exclude it.
\begin{itemize}
    \item {\bf J120838.02-004802.0}: in this case, the GAMA spectrum confirms the HSC-SSP redshift, but gives emission lines more likely associated to star-forming galaxies than AGN, i.e. $\log(\rm[OIII]/H\beta)=0.833$ and $\log(\rm[NII]/H\alpha)=-1.374$ \citep[according to the BPT diagram by][]{baldwin81}, and yields a stellar mass estimate \citep{taylor11} of about $M_{\rm star}=10^{9.6}\msun$. In addition, there is a massive galaxy ($M_{\rm star} \sim 6\times 10^{10}\msun$ at $z_{\rm best}\sim 1$ at $\sim 5.5$\arcsec\ distance in the HSC-SSP map, for which infrared emission likely overlaps;
    \item {\bf J141913.94-005952.6}: in this case, the HSC-SSP catalogue identifies several sources, all overlapping with (or next to) a strongly perturbed galaxy in foreground. The GAMA spectrum is also in this case consistent with typical star-forming galaxies, i.e. $\log(\rm[OIII]/H\beta)=0.228$ and $\log(\rm[NII]/H\alpha)=-0.6704$.
\end{itemize}

\begin{figure*}
    \centering
    \begin{tabular}{c@{\hskip 0cm}c@{\hskip 0.2cm}c@{\hskip 0cm}c}
    
    \multicolumn{2}{c}{J015948.97-055720.9}& \multicolumn{2}{c}{J021101.48-040137.0}\\
    
    \includegraphics[height=3.5cm]{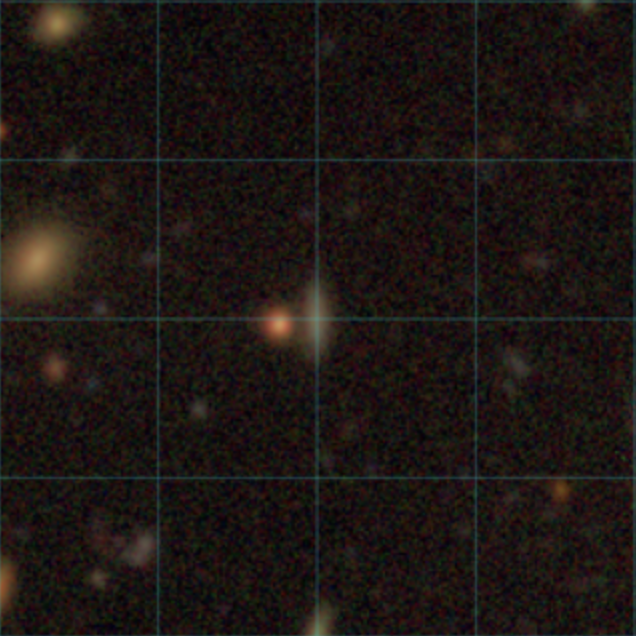}&
    \includegraphics[height=3.5cm]{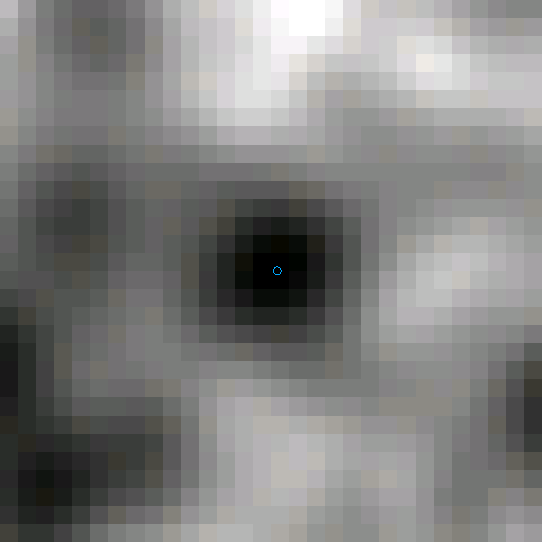}&
    \includegraphics[height=3.5cm]{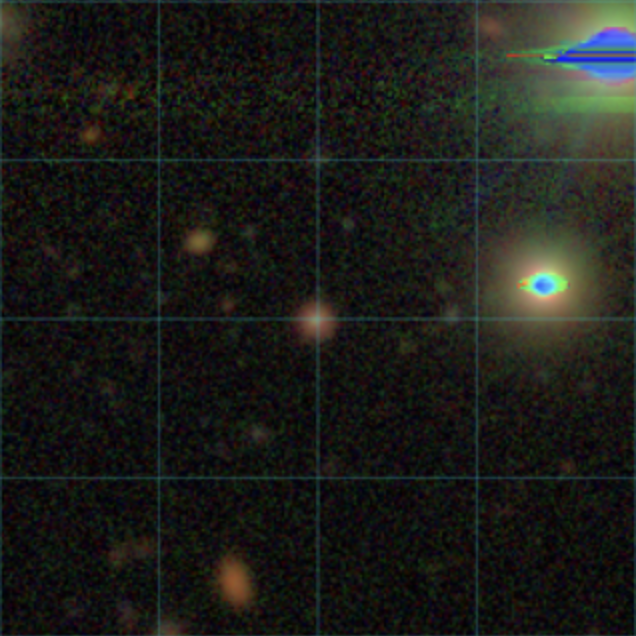}&
    \includegraphics[height=3.5cm]{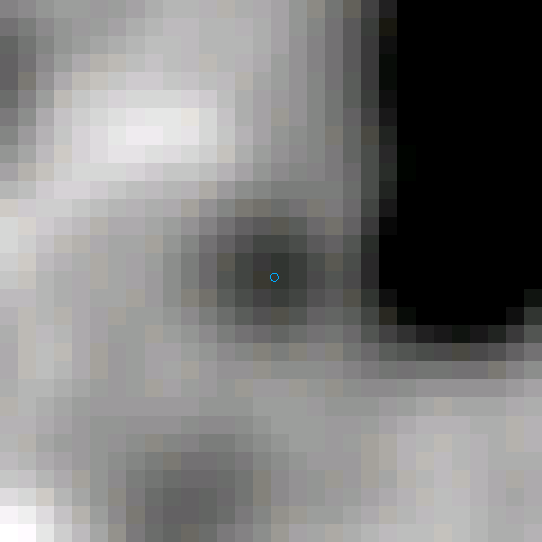}\\
    
    \multicolumn{2}{c}{J084429.75+014541.7} & \multicolumn{2}{c}{J085831.57+015648.7}\\
    \includegraphics[height=3.5cm]{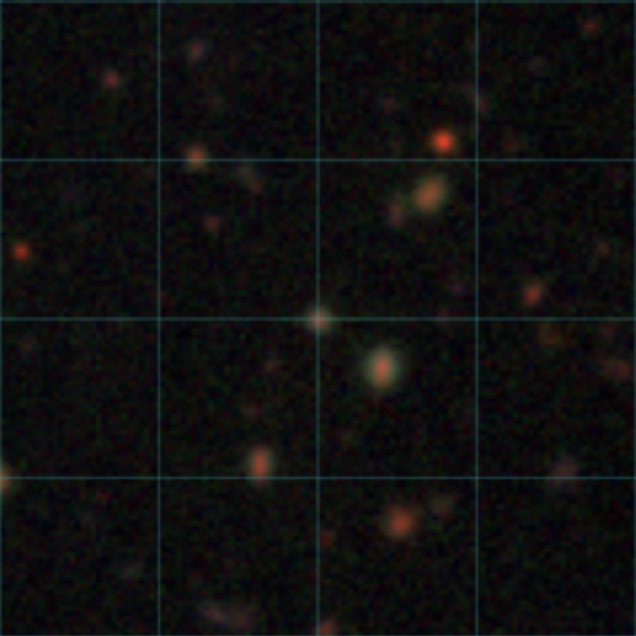}&
    \includegraphics[height=3.5cm]{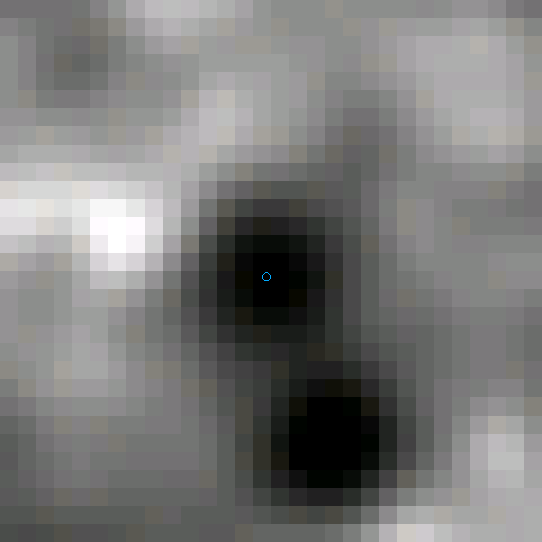}&
    \includegraphics[height=3.5cm]{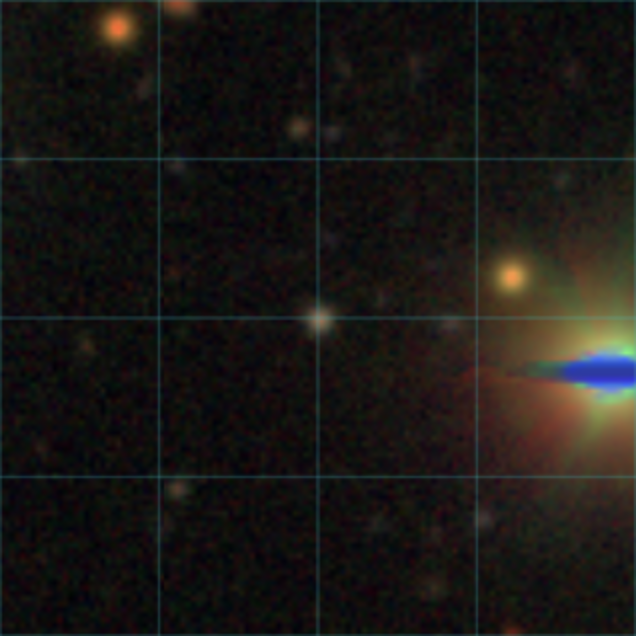}&
    \includegraphics[height=3.5cm]{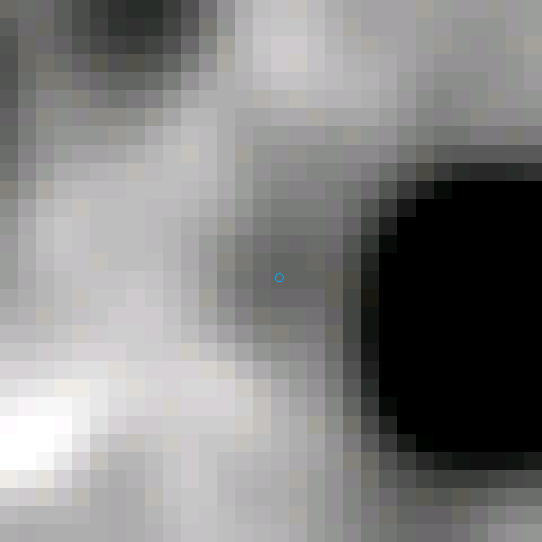}\\
    
    \multicolumn{2}{c}{J120730.58+003447.6} & \multicolumn{2}{c}{J141544.25-000953.1}\\
    \includegraphics[height=3.5cm]{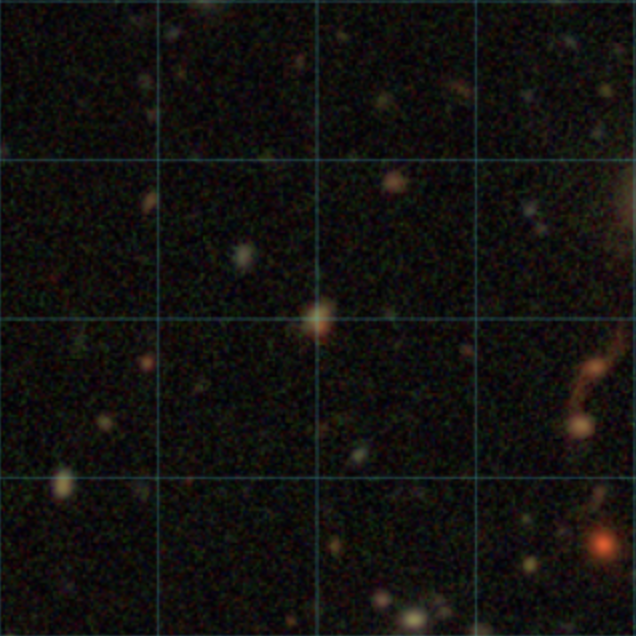}&
    \includegraphics[height=3.5cm]{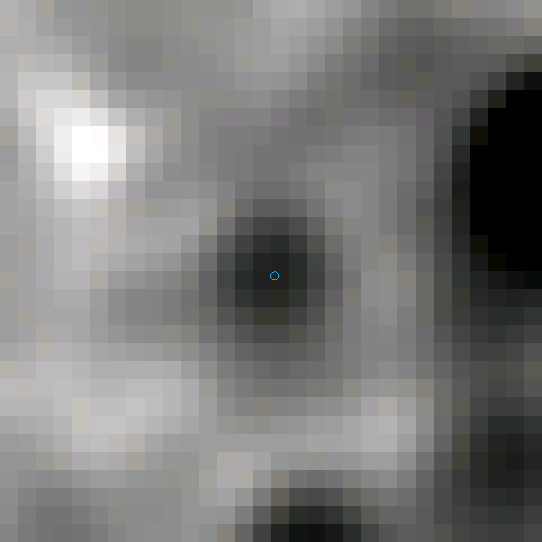}&
    \includegraphics[height=3.5cm]{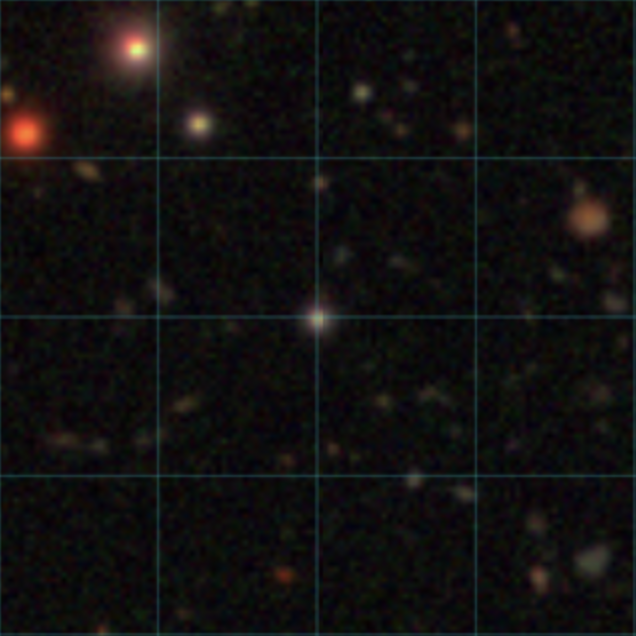}&
    \includegraphics[height=3.5cm]{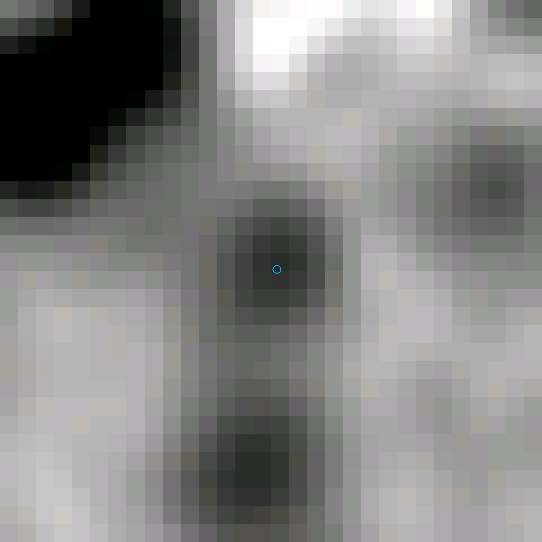}\\
    
    \multicolumn{2}{c}{J143753.75+001448.7} & \multicolumn{2}{c}{J163148.98+425240.4}\\
    \includegraphics[height=3.5cm]{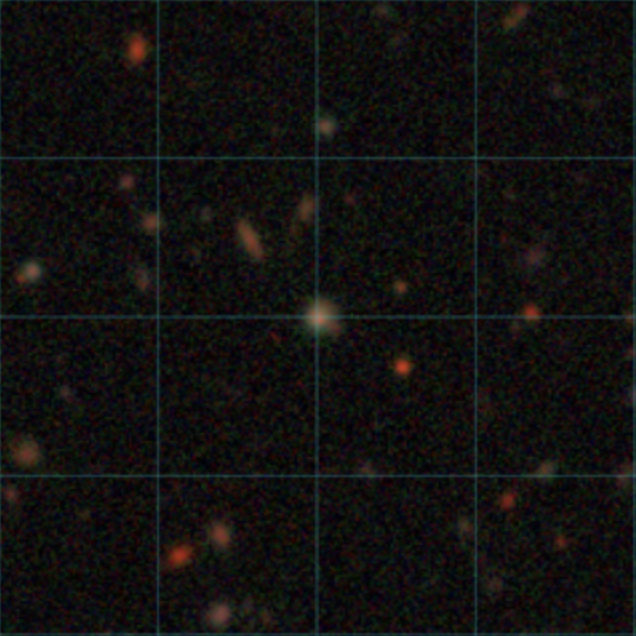}&
    \includegraphics[height=3.5cm]{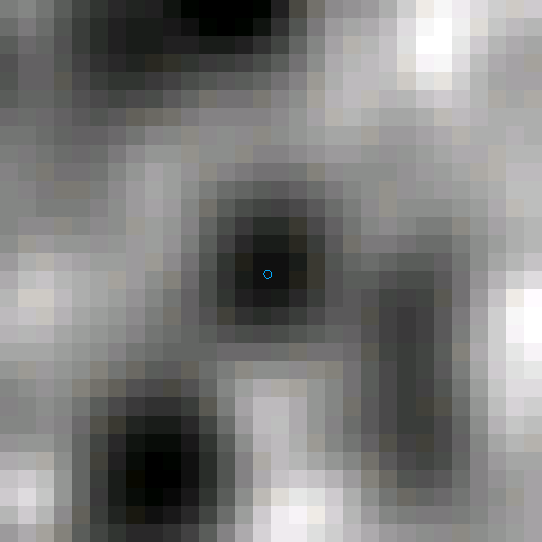}&
    \includegraphics[height=3.5cm]{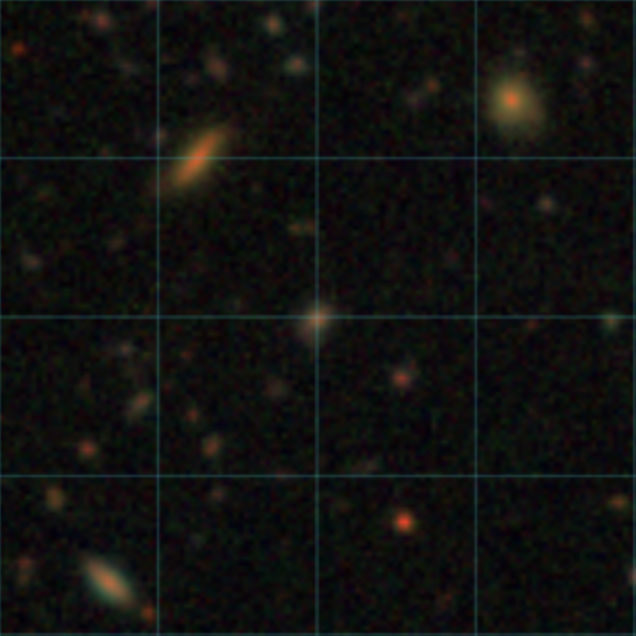}&
    \includegraphics[height=3.5cm]{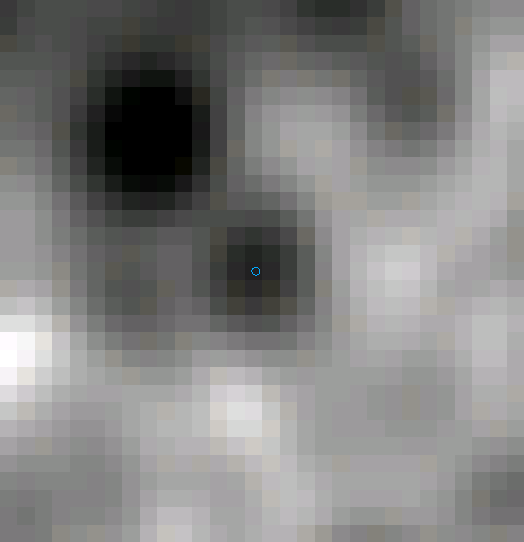}\\
    
    \multicolumn{2}{c}{J222705.43+005101.4} & \multicolumn{2}{c}{J222752.79+004431.1}\\
    \includegraphics[height=3.5cm]{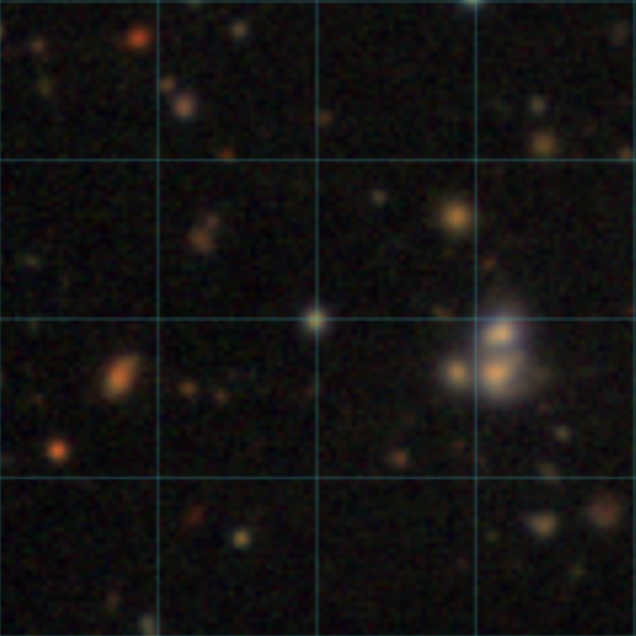}&
    \includegraphics[height=3.5cm]{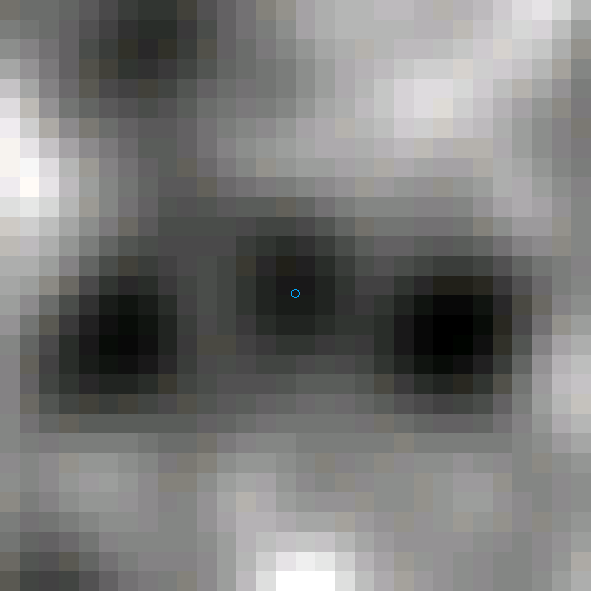}&
    \includegraphics[height=3.5cm]{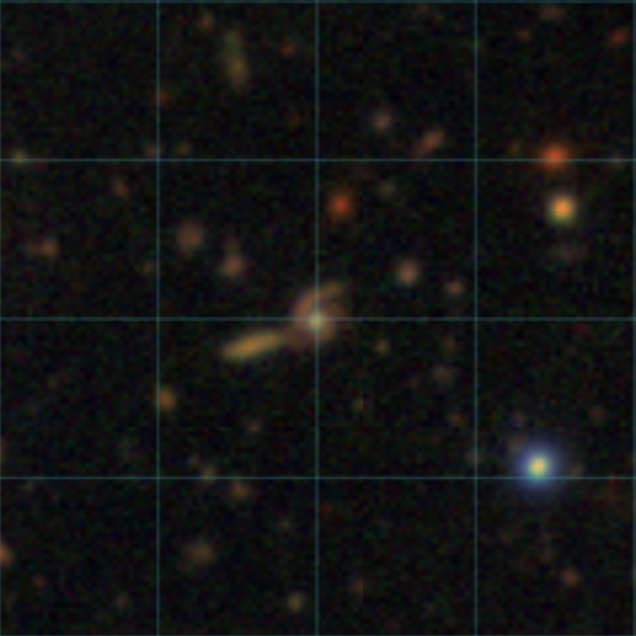}&
    \includegraphics[height=3.5cm]{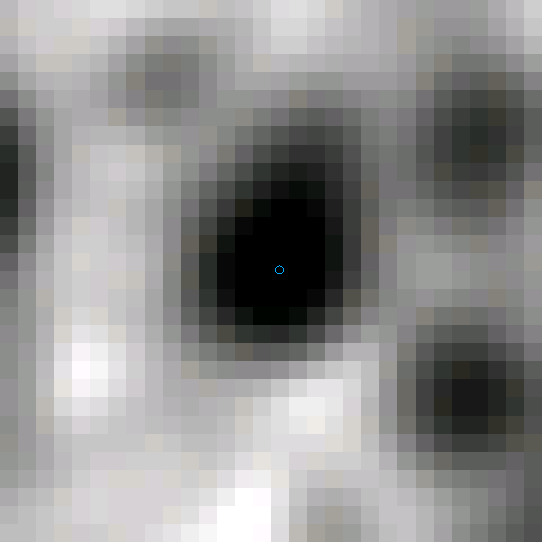}\\
    \end{tabular}
    \caption{HSC-SSP WISE composite maps of the ten best candidates in our sample. For each pair, the left-hand map corresponds to the {\it irg} map of HSC-SSP and the right-hand one to the infrared emission from WISE W1, with the maps covering a 40x40$\arcsec$ field of view. The huge different in resolution between the two surveys is striking, and a great care in the cross-matching is required.}
    \label{fig:candidates}
\end{figure*}

For the remaining objects not reported here, we can finally confirm that they represent dwarf AGN candidates from the HSC-SSP+Wise catalogue. The corresponding HSC-SSP (composite) and Wise (W1) maps, centred at the location of the target and covering a 40x40$\arcsec$ area, are reported in Fig.~\ref{fig:candidates}.

\section{Discussion and caveats}
\label{sec:discussion}
Our detailed analysis has allowed us to more accurately identify  AGN candidates in dwarf galaxies than K19. The population of likely dwarf systems, which actually exhibits strong emission in the W3 band, is represented by only 20 objects, that become 2 when we limit secure detections to S/N$_{\rm W3}>5$ and remove spurious objects for which a spectroscopic identification in other surveys is available. 

Unfortunately, relative to the positive conclusion in K19, our study shows typical bolometric luminosities for dwarf AGN around $L_{\rm bol}\lesssim 10^{44}\rm\, erg\, s^{-1}$, corresponding to MBH masses of about $\sim 10^6\msun$, consistent with expectations of dwarf galaxies harbouring low-mass MBH that struggle to grow. We notice however that the masses we report are derived under the assumption of an Eddington-accreting MBH, which is not necessarily true.

The presented analysis has some caveats that should be mentioned. 
First, our exclusion criteria for dwarf in groups including a massive system could bias the analysis, in the case the AGN is actually in the dwarf and not in the massive galaxy, albeit less likely. 
Secondly, the emission in the W3 band used to identify the AGN could also arise from different systems in the group, each accounting for a fraction of the total emission detected.  In this study, we did not consider this possible effect, but we note that to disentangle each contribution we would need higher resolution infrared observations of the same regions, currently not available.

Finally, our assumption of a quasar spectrum for dwarf AGN is possibly overestimating the bolometric luminosity, hence the MBH mass, since it does not take into account the host contribution due to ongoing star formation/starbursts, that are more common in dwarf galaxies \citep{hainline16}. However, this effect could be compensated by the fact that our MBH mass is derived assuming an Eddington-accreting MBH, which could not be the case if the gas density in the system is low. In this case, our MBH estimates should be taken as lower limits to the mass.

\section{Conclusions}
\label{sec:conclusions}
In this work, that has been mainly motivated by the strongly positive results by K19 about the role of AGN in dwarf galaxies,
we have discussed the difficulties of properly selecting a multi-wavelength sample of dwarf galaxies, assessing the presence of AGN in their cores and studying their stellar-BH relation. 

By testing different methods to cross--correlate photometric surveys with different resolutions, we found out that a careful evaluation of companions or close sources contamination is mandatory in the lower resolution survey. 
The issue is more pronounced when particularly faint, distant or small sources are the main focus of the study.
In our case, {\it WISE} could not resolve the different sources identified by HSC-SSP, and the IR luminosity derived from {\it WISE} data is clearly contaminated by various components, often missing a detailed reconstruction of dwarf galaxies IR emission.

Isolating a dwarf galaxy contribution in an IR survey is not the only delicate point: in these sources, strong contamination by non--AGN components have been observed in mid--IR colours \citep[see e.g.][]{hainline16,satyapal18}. Starburst activity may be responsible for intense mid-IR emission in many dwarf galaxies, and evaluating its actual contribution in specific objects is not straightforward.
An AGN--starburst ratio in mid--IR colours different from the average AGN might significantly affect the expected flux ratios at different wavelengths. 
In fact, scaling relations between different wavelength ranges are commonly used to derive physical parameters of standard AGN ($M_{\rm BH} \sim 10^8 \rm M_\odot$), but extending them to the extremes of the massive black hole mass function is not straightforward. Some caveats must be taken into account: the different contribution of a starburst component in dwarf galaxies might affect the standard IR--to--optical relation, but also different typical temperatures of the accretion flow on a small ($10^6\rm M_\odot$) black hole might change the typical emitting features of the host nuclear region \citep[for examples of close--to--Eddington emission in different mass ranges see][]{kubota19}. 

In the light of our results, the K19 claims about the easy detectability of AGN in dwarf galaxies, and the strong impact of AGN in these systems, seem a bit too optimistic, and probably far from the reality.
Nonetheless, after an accurate sample selection, the presence of active massive black holes in dwarf nuclei is an extremely interesting issue. For this reason, after a careful selection in IR and optical surveys, and taking into account all possible issues and caveats, we focus on a limited sample of AGN candidates in dwarf galaxies, for which spectroscopic follow--ups will be necessary. 
In particular, J222752.79+004431.1 and J084429.75+014541.7 are the only candidates with $\rm S/N_{\rm W3}>5$, making them the most promising sources in our sample.  

\section*{Acknowledgements}
AL and SC acknowledge support by the European Research Council No. 740120 `INTERSTELLAR'. \\The analysis of the catalogues reported in this work has been performed with \textsc{topcat} \citep{taylor05}.




\bibliographystyle{mnras}
\bibliography{Biblio} 

\begin{thebibliography}{}
\makeatletter
\relax
\def\mn@urlcharsother{\let\do\@makeother \do\$\do\&\do\#\do\^\do\_\do\%\do\~}
\def\mn@doi{\begingroup\mn@urlcharsother \@ifnextchar [ {\mn@doi@}
  {\mn@doi@[]}}
\def\mn@doi@[#1]#2{\def\@tempa{#1}\ifx\@tempa\@empty \href
  {http://dx.doi.org/#2} {doi:#2}\else \href {http://dx.doi.org/#2} {#1}\fi
  \endgroup}
\def\mn@eprint#1#2{\mn@eprint@#1:#2::\@nil}
\def\mn@eprint@arXiv#1{\href {http://arxiv.org/abs/#1} {{\tt arXiv:#1}}}
\def\mn@eprint@dblp#1{\href {http://dblp.uni-trier.de/rec/bibtex/#1.xml}
  {dblp:#1}}
\def\mn@eprint@#1:#2:#3:#4\@nil{\def\@tempa {#1}\def\@tempb {#2}\def\@tempc
  {#3}\ifx \@tempc \@empty \let \@tempc \@tempb \let \@tempb \@tempa \fi \ifx
  \@tempb \@empty \def\@tempb {arXiv}\fi \@ifundefined
  {mn@eprint@\@tempb}{\@tempb:\@tempc}{\expandafter \expandafter \csname
  mn@eprint@\@tempb\endcsname \expandafter{\@tempc}}}

\bibitem[\protect\citeauthoryear{{Aihara} et~al.,}{{Aihara}
  et~al.}{2018a}]{aihara18a}
{Aihara} H.,  et~al., 2018a, \mn@doi [\pasj] {10.1093/pasj/psx066}, \href
  {https://ui.adsabs.harvard.edu/abs/2018PASJ...70S...4A} {70, S4}

\bibitem[\protect\citeauthoryear{{Aihara} et~al.,}{{Aihara}
  et~al.}{2018b}]{aihara18b}
{Aihara} H.,  et~al., 2018b, \mn@doi [\pasj] {10.1093/pasj/psx081}, \href
  {https://ui.adsabs.harvard.edu/abs/2018PASJ...70S...8A} {70, S8}

\bibitem[\protect\citeauthoryear{{Angl{\'e}s-Alc{\'a}zar},
  {Faucher-Gigu{\`e}re}, {Quataert}, {Hopkins}, {Feldmann}, {Torrey}, {Wetzel}
  \& {Kere{\v s}}}{{Angl{\'e}s-Alc{\'a}zar} et~al.}{2017}]{anglesalcazar17FIRE}
{Angl{\'e}s-Alc{\'a}zar} D.,  {Faucher-Gigu{\`e}re} C.-A.,  {Quataert} E.,
  {Hopkins} P.~F.,  {Feldmann} R.,  {Torrey} P.,  {Wetzel} A.,   {Kere{\v s}}
  D.,  2017, \mn@doi [\mnras] {10.1093/mnrasl/slx161}, \href
  {http://adsabs.harvard.edu/abs/2017MNRAS.472L.109A} {472, L109}

\bibitem[\protect\citeauthoryear{{Baldwin}, {Phillips}  \&
  {Terlevich}}{{Baldwin} et~al.}{1981}]{baldwin81}
{Baldwin} J.~A.,  {Phillips} M.~M.,   {Terlevich} R.,  1981, \mn@doi [\pasp]
  {10.1086/130766}, \href
  {https://ui.adsabs.harvard.edu/abs/1981PASP...93....5B} {93, 5}

\bibitem[\protect\citeauthoryear{{Chilingarian}, {Melchior}  \&
  {Zolotukhin}}{{Chilingarian} et~al.}{2010}]{chilingarian10}
{Chilingarian} I.~V.,  {Melchior} A.-L.,   {Zolotukhin} I.~Y.,  2010, \mn@doi
  [\mnras] {10.1111/j.1365-2966.2010.16506.x}, \href
  {https://ui.adsabs.harvard.edu/abs/2010MNRAS.405.1409C} {405, 1409}

\bibitem[\protect\citeauthoryear{{Croom}, {Smith}, {Boyle}, {Shanks}, {Miller},
  {Outram}  \& {Loaring}}{{Croom} et~al.}{2004}]{croom04}
{Croom} S.~M.,  {Smith} R.~J.,  {Boyle} B.~J.,  {Shanks} T.,  {Miller} L.,
  {Outram} P.~J.,   {Loaring} N.~S.,  2004, \mn@doi [\mnras]
  {10.1111/j.1365-2966.2004.07619.x}, \href
  {https://ui.adsabs.harvard.edu/abs/2004MNRAS.349.1397C} {349, 1397}

\bibitem[\protect\citeauthoryear{{Di Matteo}, {Springel}  \& {Hernquist}}{{Di
  Matteo} et~al.}{2005}]{dimatteo05}
{Di Matteo} T.,  {Springel} V.,   {Hernquist} L.,  2005, \mn@doi [\nat]
  {10.1038/nature03335}, \href
  {http://adsabs.harvard.edu/abs/2005Natur.433..604D} {433, 604}

\bibitem[\protect\citeauthoryear{{Driver} et~al.,}{{Driver}
  et~al.}{2009}]{driver09}
{Driver} S.~P.,  et~al., 2009, \mn@doi [Astronomy and Geophysics]
  {10.1111/j.1468-4004.2009.50512.x}, \href
  {https://ui.adsabs.harvard.edu/abs/2009A%26G....50e..12D} {50, 5.12}

\bibitem[\protect\citeauthoryear{{Dubois} et~al.,}{{Dubois}
  et~al.}{2014}]{dubois14}
{Dubois} Y.,  et~al., 2014, \mn@doi [\mnras] {10.1093/mnras/stu1227}, \href
  {http://adsabs.harvard.edu/abs/2014MNRAS.444.1453D} {444, 1453}

\bibitem[\protect\citeauthoryear{{Ferrarese} \& {Merritt}}{{Ferrarese} \&
  {Merritt}}{2000}]{ferrarese00}
{Ferrarese} L.,  {Merritt} D.,  2000, \mn@doi [\apjl] {10.1086/312838}, \href
  {http://adsabs.harvard.edu/abs/2000ApJ...539L...9F} {539, L9}

\bibitem[\protect\citeauthoryear{{Garrison-Kimmel}, {Rocha}, {Boylan-Kolchin},
  {Bullock}  \& {Lally}}{{Garrison-Kimmel} et~al.}{2013}]{garrisonkimmel13}
{Garrison-Kimmel} S.,  {Rocha} M.,  {Boylan-Kolchin} M.,  {Bullock} J.~S.,
  {Lally} J.,  2013, \mn@doi [\mnras] {10.1093/mnras/stt984}, \href
  {https://ui.adsabs.harvard.edu/abs/2013MNRAS.433.3539G} {433, 3539}

\bibitem[\protect\citeauthoryear{{Greene}}{{Greene}}{2012}]{greene12}
{Greene} J.~E.,  2012, \mn@doi [Nature Communications] {10.1038/ncomms2314},
  \href {https://ui.adsabs.harvard.edu/abs/2012NatCo...3.1304G} {3, 1304}

\bibitem[\protect\citeauthoryear{{Greene}, {Strader}  \& {Ho}}{{Greene}
  et~al.}{2019}]{greene19}
{Greene} J.~E.,  {Strader} J.,   {Ho} L.~C.,  2019, arXiv e-prints, \href
  {https://ui.adsabs.harvard.edu/abs/2019arXiv191109678G} {p. arXiv:1911.09678}

\bibitem[\protect\citeauthoryear{{G{\"u}ltekin} et~al.,}{{G{\"u}ltekin}
  et~al.}{2009}]{gultekin09}
{G{\"u}ltekin} K.,  et~al., 2009, \mn@doi [\apj] {10.1088/0004-637X/698/1/198},
  \href {http://adsabs.harvard.edu/abs/2009ApJ...698..198G} {698, 198}

\bibitem[\protect\citeauthoryear{{Habouzit}, {Volonteri}  \&
  {Dubois}}{{Habouzit} et~al.}{2017}]{habouzit17}
{Habouzit} M.,  {Volonteri} M.,   {Dubois} Y.,  2017, \mn@doi [\mnras]
  {10.1093/mnras/stx666}, \href
  {http://adsabs.harvard.edu/abs/2017MNRAS.468.3935H} {468, 3935}

\bibitem[\protect\citeauthoryear{{Hainline}, {Reines}, {Greene}  \&
  {Stern}}{{Hainline} et~al.}{2016}]{hainline16}
{Hainline} K.~N.,  {Reines} A.~E.,  {Greene} J.~E.,   {Stern} D.,  2016,
  \mn@doi [\apj] {10.3847/0004-637X/832/2/119}, \href
  {https://ui.adsabs.harvard.edu/abs/2016ApJ...832..119H} {832, 119}

\bibitem[\protect\citeauthoryear{{Hopkins} et~al.,}{{Hopkins}
  et~al.}{2013}]{hopkins13gama}
{Hopkins} A.~M.,  et~al., 2013, \mn@doi [\mnras] {10.1093/mnras/stt030}, \href
  {https://ui.adsabs.harvard.edu/abs/2013MNRAS.430.2047H} {430, 2047}

\bibitem[\protect\citeauthoryear{{Jarrett} et~al.,}{{Jarrett}
  et~al.}{2011}]{jarrett11}
{Jarrett} T.~H.,  et~al., 2011, \mn@doi [\apj] {10.1088/0004-637X/735/2/112},
  \href {https://ui.adsabs.harvard.edu/abs/2011ApJ...735..112J} {735, 112}

\bibitem[\protect\citeauthoryear{{Kaviraj}, {Martin}  \& {Silk}}{{Kaviraj}
  et~al.}{2019}]{kaviraj19}
{Kaviraj} S.,  {Martin} G.,   {Silk} J.,  2019, \mn@doi [\mnras]
  {10.1093/mnrasl/slz102}, \href
  {https://ui.adsabs.harvard.edu/abs/2019MNRAS.489L..12K} {489, L12}

\bibitem[\protect\citeauthoryear{{Kormendy} \& {Ho}}{{Kormendy} \&
  {Ho}}{2013}]{kormendy13bh}
{Kormendy} J.,  {Ho} L.~C.,  2013, \mn@doi [\araa]
  {10.1146/annurev-astro-082708-101811}, \href
  {http://adsabs.harvard.edu/abs/2013ARA%26A..51..511K} {51, 511}

\bibitem[\protect\citeauthoryear{{Kubota} \& {Done}}{{Kubota} \&
  {Done}}{2019}]{kubota19}
{Kubota} A.,  {Done} C.,  2019, \mn@doi [\mnras] {10.1093/mnras/stz2140}, \href
  {https://ui.adsabs.harvard.edu/abs/2019MNRAS.489..524K} {489, 524}

\bibitem[\protect\citeauthoryear{{Mackay Dickey}, {Geha}, {Wetzel}  \&
  {El-Badry}}{{Mackay Dickey} et~al.}{2019}]{mackaydickey19}
{Mackay Dickey} C.,  {Geha} M.,  {Wetzel} A.,   {El-Badry} K.,  2019, \mn@doi
  [\apj] {10.3847/1538-4357/ab3220}, \href
  {https://ui.adsabs.harvard.edu/abs/2019ApJ...884..180M} {884, 180}

\bibitem[\protect\citeauthoryear{{Manzano-King}, {Canalizo}  \&
  {Sales}}{{Manzano-King} et~al.}{2019}]{manzanoking19}
{Manzano-King} C.~M.,  {Canalizo} G.,   {Sales} L.~V.,  2019, \mn@doi [\apj]
  {10.3847/1538-4357/ab4197}, \href
  {https://ui.adsabs.harvard.edu/abs/2019ApJ...884...54M} {884, 54}

\bibitem[\protect\citeauthoryear{{Marleau}, {Clancy}, {Habas}  \&
  {Bianconi}}{{Marleau} et~al.}{2017}]{marleau17}
{Marleau} F.~R.,  {Clancy} D.,  {Habas} R.,   {Bianconi} M.,  2017, \mn@doi
  [\aap] {10.1051/0004-6361/201629832}, \href
  {https://ui.adsabs.harvard.edu/abs/2017A&A...602A..28M} {602, A28}

\bibitem[\protect\citeauthoryear{{Mezcua}, {Civano}, {Fabbiano}, {Miyaji}  \&
  {Marchesi}}{{Mezcua} et~al.}{2016}]{mezcua16}
{Mezcua} M.,  {Civano} F.,  {Fabbiano} G.,  {Miyaji} T.,   {Marchesi} S.,
  2016, \mn@doi [\apj] {10.3847/0004-637X/817/1/20}, \href
  {https://ui.adsabs.harvard.edu/abs/2016ApJ...817...20M} {817, 20}

\bibitem[\protect\citeauthoryear{{Mezcua}, {Civano}, {Marchesi}, {Suh},
  {Fabbiano}  \& {Volonteri}}{{Mezcua} et~al.}{2018}]{mezcua18}
{Mezcua} M.,  {Civano} F.,  {Marchesi} S.,  {Suh} H.,  {Fabbiano} G.,
  {Volonteri} M.,  2018, \mn@doi [\mnras] {10.1093/mnras/sty1163}, \href
  {https://ui.adsabs.harvard.edu/abs/2018MNRAS.478.2576M} {478, 2576}

\bibitem[\protect\citeauthoryear{{Miller}, {Gallo}, {Greene}, {Kelly}, {Treu},
  {Woo}  \& {Baldassare}}{{Miller} et~al.}{2015}]{miller15}
{Miller} B.~P.,  {Gallo} E.,  {Greene} J.~E.,  {Kelly} B.~C.,  {Treu} T.,
  {Woo} J.-H.,   {Baldassare} V.,  2015, \mn@doi [\apj]
  {10.1088/0004-637X/799/1/98}, \href
  {https://ui.adsabs.harvard.edu/abs/2015ApJ...799...98M} {799, 98}

\bibitem[\protect\citeauthoryear{{Peirani}, {Jung}, {Silk}  \&
  {Pichon}}{{Peirani} et~al.}{2012}]{peirani12}
{Peirani} S.,  {Jung} I.,  {Silk} J.,   {Pichon} C.,  2012, \mn@doi [\mnras]
  {10.1111/j.1365-2966.2012.22105.x}, \href
  {https://ui.adsabs.harvard.edu/abs/2012MNRAS.427.2625P} {427, 2625}

\bibitem[\protect\citeauthoryear{{Penny} et~al.,}{{Penny}
  et~al.}{2018}]{penny18}
{Penny} S.~J.,  et~al., 2018, \mn@doi [\mnras] {10.1093/mnras/sty202}, \href
  {https://ui.adsabs.harvard.edu/abs/2018MNRAS.476..979P} {476, 979}

\bibitem[\protect\citeauthoryear{{Prieto}, {Escala}, {Volonteri}  \&
  {Dubois}}{{Prieto} et~al.}{2017}]{prieto17}
{Prieto} J.,  {Escala} A.,  {Volonteri} M.,   {Dubois} Y.,  2017, \mn@doi
  [\apj] {10.3847/1538-4357/aa5be5}, \href
  {http://adsabs.harvard.edu/abs/2017ApJ...836..216P} {836, 216}

\bibitem[\protect\citeauthoryear{{Reines} \& {Comastri}}{{Reines} \&
  {Comastri}}{2016}]{reines16}
{Reines} A.~E.,  {Comastri} A.,  2016, \mn@doi [\pasa] {10.1017/pasa.2016.46},
  \href {https://ui.adsabs.harvard.edu/abs/2016PASA...33...54R} {33, e054}

\bibitem[\protect\citeauthoryear{{Reines} \& {Volonteri}}{{Reines} \&
  {Volonteri}}{2015}]{reines15}
{Reines} A.~E.,  {Volonteri} M.,  2015, \mn@doi [\apj]
  {10.1088/0004-637X/813/2/82}, \href
  {http://adsabs.harvard.edu/abs/2015ApJ...813...82R} {813, 82}

\bibitem[\protect\citeauthoryear{{Reines}, {Greene}  \& {Geha}}{{Reines}
  et~al.}{2013}]{reines13}
{Reines} A.~E.,  {Greene} J.~E.,   {Geha} M.,  2013, \mn@doi [\apj]
  {10.1088/0004-637X/775/2/116}, \href
  {https://ui.adsabs.harvard.edu/abs/2013ApJ...775..116R} {775, 116}

\bibitem[\protect\citeauthoryear{{Reines}, {Condon}, {Darling}  \&
  {Greene}}{{Reines} et~al.}{2019}]{reines19}
{Reines} A.,  {Condon} J.,  {Darling} J.,   {Greene} J.,  2019, arXiv e-prints,
  \href {https://ui.adsabs.harvard.edu/abs/2019arXiv190904670R} {p.
  arXiv:1909.04670}

\bibitem[\protect\citeauthoryear{{Richards} et~al.,}{{Richards}
  et~al.}{2006}]{richards06}
{Richards} G.~T.,  et~al., 2006, \mn@doi [\apjs] {10.1086/506525}, \href
  {https://ui.adsabs.harvard.edu/abs/2006ApJS..166..470R} {166, 470}

\bibitem[\protect\citeauthoryear{{Sartori}, {Schawinski}, {Treister},
  {Trakhtenbrot}, {Koss}, {Shirazi}  \& {Oh}}{{Sartori}
  et~al.}{2015}]{sartori15}
{Sartori} L.~F.,  {Schawinski} K.,  {Treister} E.,  {Trakhtenbrot} B.,  {Koss}
  M.,  {Shirazi} M.,   {Oh} K.,  2015, \mn@doi [\mnras]
  {10.1093/mnras/stv2238}, \href
  {https://ui.adsabs.harvard.edu/abs/2015MNRAS.454.3722S} {454, 3722}

\bibitem[\protect\citeauthoryear{{Satyapal}, {Secrest}, {McAlpine}, {Ellison},
  {Fischer}  \& {Rosenberg}}{{Satyapal} et~al.}{2014}]{satyapal14}
{Satyapal} S.,  {Secrest} N.~J.,  {McAlpine} W.,  {Ellison} S.~L.,  {Fischer}
  J.,   {Rosenberg} J.~L.,  2014, \mn@doi [\apj] {10.1088/0004-637X/784/2/113},
  \href {https://ui.adsabs.harvard.edu/abs/2014ApJ...784..113S} {784, 113}

\bibitem[\protect\citeauthoryear{{Satyapal}, {Abel}  \& {Secrest}}{{Satyapal}
  et~al.}{2018}]{satyapal18}
{Satyapal} S.,  {Abel} N.~P.,   {Secrest} N.~J.,  2018, \mn@doi [\apj]
  {10.3847/1538-4357/aab7f8}, \href
  {https://ui.adsabs.harvard.edu/abs/2018ApJ...858...38S} {858, 38}

\bibitem[\protect\citeauthoryear{{Silk} \& {Rees}}{{Silk} \&
  {Rees}}{1998}]{silk98}
{Silk} J.,  {Rees} M.~J.,  1998, \aap, \href
  {http://adsabs.harvard.edu/abs/1998A%26A...331L...1S} {331, L1}

\bibitem[\protect\citeauthoryear{{Tanaka}}{{Tanaka}}{2015}]{tanaka15}
{Tanaka} M.,  2015, \mn@doi [\apj] {10.1088/0004-637X/801/1/20}, \href
  {https://ui.adsabs.harvard.edu/abs/2015ApJ...801...20T} {801, 20}

\bibitem[\protect\citeauthoryear{{Tanaka} et~al.,}{{Tanaka}
  et~al.}{2018}]{tanaka18}
{Tanaka} M.,  et~al., 2018, \mn@doi [\pasj] {10.1093/pasj/psx077}, \href
  {https://ui.adsabs.harvard.edu/abs/2018PASJ...70S...9T} {70, S9}

\bibitem[\protect\citeauthoryear{{Taylor}}{{Taylor}}{2005}]{taylor05}
{Taylor} M.~B.,  2005, {TOPCAT and STIL: Starlink Table/VOTable Processing
  Software}.
{Shopbell}, P. and {Britton}, M. and {Ebert}, R., p.~29

\bibitem[\protect\citeauthoryear{{Taylor} et~al.,}{{Taylor}
  et~al.}{2011}]{taylor11}
{Taylor} E.~N.,  et~al., 2011, \mn@doi [\mnras]
  {10.1111/j.1365-2966.2011.19536.x}, \href
  {https://ui.adsabs.harvard.edu/abs/2011MNRAS.418.1587T} {418, 1587}

\bibitem[\protect\citeauthoryear{{Trebitsch}, {Volonteri}, {Dubois}  \&
  {Madau}}{{Trebitsch} et~al.}{2018}]{trebitsch18}
{Trebitsch} M.,  {Volonteri} M.,  {Dubois} Y.,   {Madau} P.,  2018, \mn@doi
  [\mnras] {10.1093/mnras/sty1406}, \href
  {http://adsabs.harvard.edu/abs/2018MNRAS.478.5607T} {478, 5607}

\bibitem[\protect\citeauthoryear{{Volonteri} \& {Gnedin}}{{Volonteri} \&
  {Gnedin}}{2009}]{volonterignedin09}
{Volonteri} M.,  {Gnedin} N.~Y.,  2009, \mn@doi [\apj]
  {10.1088/0004-637X/703/2/2113}, \href
  {https://ui.adsabs.harvard.edu/abs/2009ApJ...703.2113V} {703, 2113}

\bibitem[\protect\citeauthoryear{{Wright} et~al.,}{{Wright}
  et~al.}{2010}]{wright10}
{Wright} E.~L.,  et~al., 2010, \mn@doi [\aj] {10.1088/0004-6256/140/6/1868},
  \href {https://ui.adsabs.harvard.edu/abs/2010AJ....140.1868W} {140, 1868}

\makeatother
\end{thebibliography}


\appendix
\section{The role of the S/N$_{\rm W3}$ cut}\label{app:SNcut}
\begin{figure}
    \centering
    \includegraphics[width=\columnwidth]{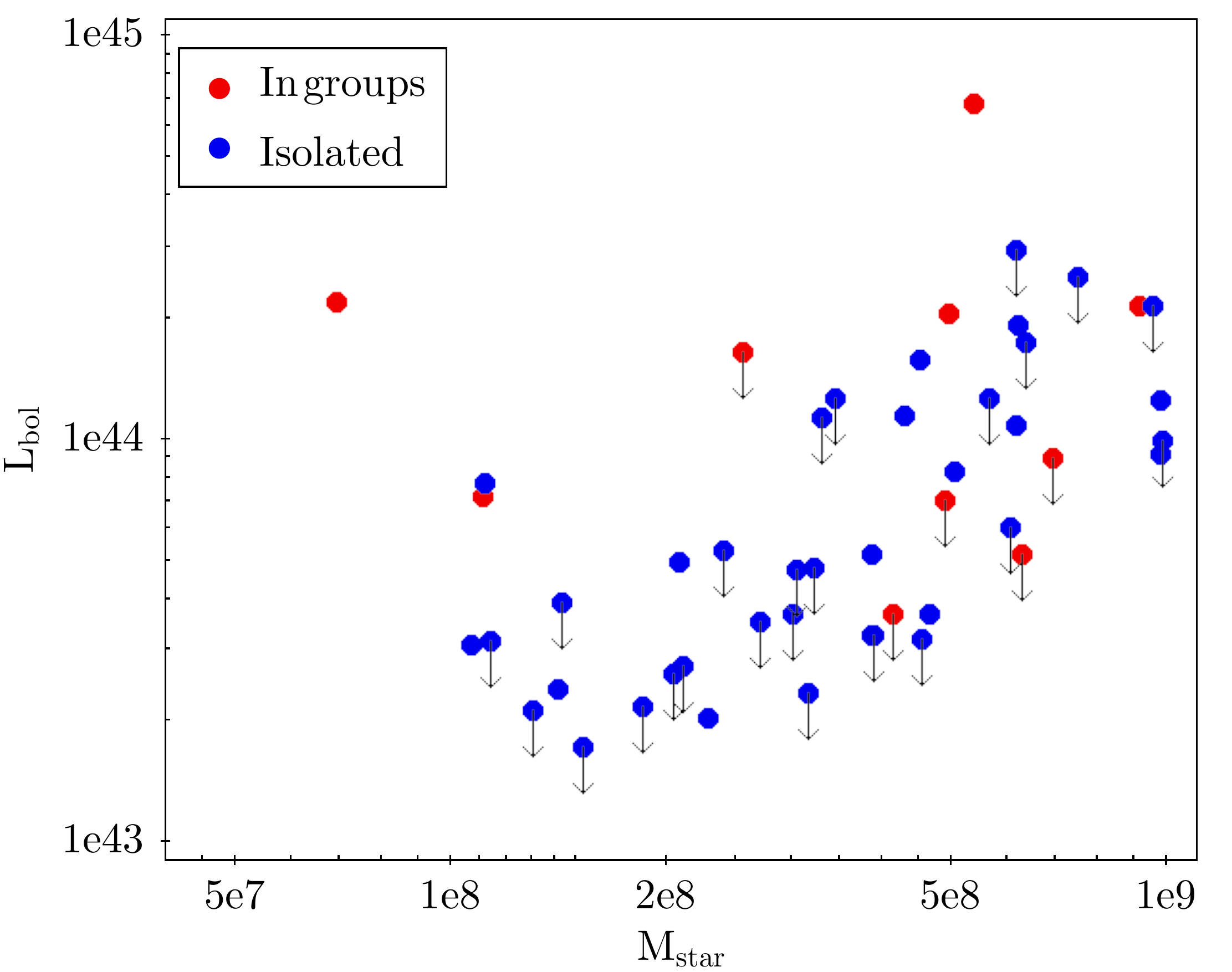}
    \caption{Same as Fig.~\ref{fig:lbol} for the extended sample, with black arrows corresponding to the objects for which only upper limits to the W3 flux are available. Also in this case, the data settle around $L_{\rm bol}=10^{44}\rm\, erg\, s^{-1}$, much lower than the $10^{46}\rm\, erg\, s^{-1}$ reported in K19, that could hardly be explained by a MBH with less than $10^8\msun$.}
    \label{fig:lboltot}
\end{figure}

Here, we show how the removal of the S/N$_{\rm W3}$ cut we applied in Section~\ref{sec:data} affects our conclusions.
When we remove the cut (that corresponds to including objects for which WISE only has upper limits to the actual flux), as done in K19, we double the number of AGN candidate in the sample, moving from $\sim 0.4$ per cent to $\sim 0.9$ per cent, but this is still much lower than the value reported in K19 of about 10 per cent. Obviously, these values only represent a lower limit to the real AGN fraction, since their identification is based on a strong mid-infrared emission. Assuming that, also for this extended sample, the W3 flux is completely associated to the dust heating by an AGN, we show in Fig.~\ref{fig:lboltot} the expected bolometric luminosity, as done in Fig.~\ref{fig:lbol}. Interestingly, also in this case we get a median bolometric luminosity of  $10^{44}$~erg s$^{-1}$, two orders of magnitude lower than that reported in K19 (the black arrows represent data for which only upper limits to the W3 flux are available).
\bsp	
\label{lastpage}
\end{document}